\documentclass[12pt]{iopart}
\usepackage[utf8]{inputenc}
\usepackage[T1]{fontenc}
\usepackage{cite}
\usepackage{gensymb}
\usepackage{amssymb}
\usepackage{eurosym}
\usepackage{graphicx}
\DeclareUnicodeCharacter{20AC}{\euro{}}
\DeclareUnicodeCharacter{2082}{$_2$}
\DeclareUnicodeCharacter{B2}{$^2$}
\usepackage{pgfkeys}
\usepackage{csvsimple}
\usepackage{xtab}
\usepackage{ifpdf}
\ifpdf
 \usepackage[pdftex,linktocpage=true,unicode=true,colorlinks=true,citecolor=red,urlcolor=blue,
 			pdfstartview=FitH,
            pdfauthor={Timur Delahaye et al.},
            pdftitle={Negative Emissions in Sweden},
            pdfsubject={Negative Emissions in Sweden},
            pdfkeywords={Bio-CCS, BECCS, COP21, 1.5\degree C target, Negative Emission Technologies.},
            pdfproducer={Latex with hyperref},
            pdfcreator=pdflatex]{hyperref}
\else
 \usepackage[linktocpage=true,unicode=true,colorlinks=true,citecolor=red,urlcolor=blue,pdfstartview=FitH]{hyperref}
\fi
\graphicspath{{Figs/},{./}}
\begin{document}

\title[Negative Emissions in Sweden]{Immediate deployment opportunities for negative emissions with BECCS: a Swedish case study}

\author{Henrik Karlsson$^1$, Timur Delahaye$^1$, Filip Johnsson$^2$,  
Jan Kjärstad$^2$ and Johan Rootzén$^2$}

\address{$^1$ Biorecro AB, Norrsken House, Stockholm, Sweden}
\address{$^2$ Department of Energy and Environment, Chalmers University of Technology, Göteborg, Sweden}
\ead{henrik.karlsson@biorecro.se}

\begin{abstract}
To meet the 2\degree C target and, in particular the 1.5\degree C target defined in the Paris Agreement, rapid scaling-up of BECCS (Bio-Energy with Carbon Capture and Storage) and other negative emissions technologies (NETs) is essential. Recent research on BECCS has mainly focused on biophysical and sustainability limitations to multi-Gigatonne deployment in the latter half of this century. However, this paper focuses on the critical short-term opportunities for immediate deployment, considering solely existing bio-energy facilities in Sweden as a case study. We show that the immediate potential for BECCS in this country amounts to 20~Mt annually. This corresponds to 39\% of total GHG emissions in 2014 in Sweden. The current costs for implementing BECCS at this level is compared to the present carbon taxes and other incentives. We show that including BECCS in the carbon tax incentive mechanism at current incentive levels would yield 16.7~Mt of negative emissions annually with an estimated societal cost saving of more than 600~M€ annually, when compared to current incentive marginal abatement costs. We conclude that Sweden is ideally positioned for immediate BECCS deployment.

\end{abstract}

%

\submitto{\ERL}
%
\maketitle
%
%

\section{Introduction}

The international community has acknowledged that climate change is a well-established phenomenon driven by anthropogenic emissions of greenhouse gases (GHG), mainly CO₂~\cite{IPCC2013}. Gathered in Paris at the COP21 meeting, heads of states from around the world committed to work on restricting the increase in global warming to well below 2\degree C and to even aim for 1.5\degree C by the end of this century. The challenge is two-fold. First, it is crucial that the combined efforts of the different countries attain the required level. To date, the so-called Intended Nationally Determined Contributions (INDC) have not proven sufficient to reach the targets set. However the Paris Agreement includes mechanisms to ensure that these commitments, now called NDC, converge towards the required levels~\cite{Spencer2015}. Second, the various countries have to deliver and actually achieve the emissions reductions that they have committed to within the promised time-frame.

The IPCC AR5 report has emphasized that high levels of negative emissions of GHG are necessary to achieve the targets set by the Paris Agreement. This is supported by recent work that advocates for the rapid deployment of Negative Emissions Technologies (NET) in parallel with emissions reductions of 50\% every decade~\cite{Rockstrom2017}. One of the most promising approaches to producing negative emissions of the required magnitude is Bio-Energy with Carbon Capture and Storage (BECCS, sometimes also referred to as BioCCS). Although the authors do not all agree as to what extent BECCS can be deployed in a sustainable way~\cite{Muratori2016b,Hansen2016}, most studies that have focused on a given geographic area confirm that it is a potent and necessary tool for achieving large cuts in emission in a sustainable manner~\cite{Moreira2016a,Kraxner2014,Kraxner2014b,Selosse2014}. It should be noted that while other NETs are also mentioned in the literature, including bioenergy-biochar systems (BEBCS)~\cite{Woolf2016}, we decided in this work to focus on BECCS, as it is the most mature NET, having already been demonstrated on the Mt scale~\cite{Gollakota2012}.

The Swedish Government recently presented a climate act, supported by six out of eight political parties, which will bind all future governments to achieving climate neutrality at the latest in 2045~\cite{TheSwedishGovernment2017}. However, the measures through which this climate neutrality will be achieved are not yet fully determined. The Swedish Energy Agency (\textit{Energimyndigheten}) has explored four scenarios for the future of the Swedish energy system~\cite{Energimyndigheten2016}. The only scenario that would achieve climate neutrality within the time-frame set by the new act is the one that makes use of negative emissions using BECCS. However the report does not explicitly quantify the amount of negative emissions required.

The purpose of the present work is to study the potential in Sweden for BECCS  deployment in the short term and at a competitive cost relative to current marginal abatement costs in Sweden. In what follows, we describe the methods used for identifying facilities that are suitable for BECCS in Sweden, as well as for estimating the costs for CO₂ capture, transport, and storage. After presenting our results, we will compare them to current carbon pricing in Sweden and draw key conclusions.

\section{Methodology}
\subsection{Selection of biogenic emitters}
\label{subsec:selection}
In this article, we make use of data available from the European Pollutant Release and Transfer Register (E-PRTR) of the European Environmental Agency\footnote{A permalink to the version used here (v10) is~\url{https://goo.gl/CsjwRb}}, which contains information on all emissions from the European Union and EFTA countries for Year 2014. A more detailed description of the E-PRTR can be found in~\ref{app:prtr}.

\begin{figure}[ht]
\begin{center}
\includegraphics[width=0.6\textwidth]{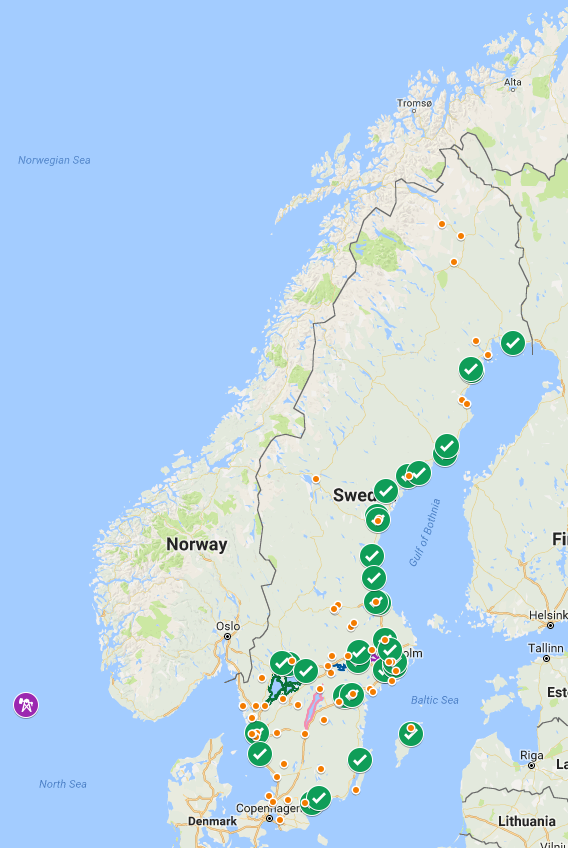}
\caption{This map shows the 96 Swedish facilities that emit more than 100 kt/y of CO₂, as reported in the E-PRTR in 2014 (in red and green). In green, are those facilities that are located less than 25~km from the sea, Lake Vänern or Lake Mälaren and have a total emissions level >300 kt/y and non-zero biogenic CO₂ emissions.
Data were extracted from the E-PRTR v10. This map is also available online: \url{https://goo.gl/seVfkM} }
\label{fig:map}
\end{center}
\end{figure}

The dataset covers 96 facilities in Sweden, including their total and biogenic CO₂ emissions, their locations, and their main activities. Using the method described in~\ref{App:distances}, we estimated for each of these facilities, the distances to the sea and to lakes Vänern and Mälaren. This distance ranges from 0 to 167~km (the upper value corresponding to the mining operations of Gällivare). This dataset is shown on the map in Figure~\ref{fig:map}, which is also available online (\url{https://goo.gl/seVfkM}).

These 96 facilities collectively emitted 48~Mt CO₂ in 2014, of which 31~Mt was of biogenic origin. These levels are comparable to the fossil CO₂ emissions reported by Sweden for the same Year 2014 (42~Mt) and the total GHG emissions (52~MtCO$_{2e}$ see \url{https://goo.gl/c61Ofb}). A list of these facilities can be found in Table~\ref{tab:data}.

However, not all of these facilities are suitable for the rapid deployment of BECCS, as either they are too far from the sea (which increases considerably the costs for transportation of the CO₂) or their total CO₂ emissions are too low. We have selected only those facilities that fulfill the following three criteria:
\begin{itemize}
\item Distance to the sea (or lakes Vänern or Mälaren) is less than \textbf{25~km}
\item Total CO₂ emissions greater than \textbf{300 kt/y}
\item  A non-zero share of biogenic CO₂ emissions.
\end{itemize}

\subsection{Capture costs}
\label{subsec:capture}
Estimating the cost of installing capture technologies at each individual plant is beyond the scope of the present work. Indeed, this cost depends on the precise configuration of the facility, the number of boilers or clinkers in the facility, and the capture technology chosen, as well as whether or not the facility sells its waste heat and steam. As such a detailed analysis is not required here, we decided to assign a reference capture cost for each category of activity, as well as a cost range that reflects the current state-of-the-art knowledge found in the literature.

The main activity of each facility is available in the E-PRTR in two forms: the NACE code and IA sector. For most cases, these two pieces of information concord. However, in some cases, the fact that the facility has several activities can lead to conflicting information (see~\ref{app:prtr} for details). Therefore, we decided to regroup the facilities into four categories following the method of~\cite{Arasto2014,Onarheim2015}. We adopted the capture costs listed in see Table~\ref{tab:captcost}.

\begin{table}[ht]
\caption{\label{tab:captcost}Costs related to CO₂ capture.}
\begin{indented}
\item[]\begin{tabular}{@{}lcc}
\br
Activity & Reference Cost [€/t] & Cost range [€/t] \\
\mr
Cement \& lime& 75 &40-110 \cite{Koring2013} \\
Pulp \& paper & 40 &16-62~\cite{Hektor2009} \\
Power & 62~\cite{Carbo2011} & 29.9-109.7\cite{ZEP2011c}\\
Ethanol fermentation & 25 & 18 \cite{Laude2011} - 40 \cite{McLaren2012}\\
\br
\end{tabular}
\end{indented}
\end{table}

The so-called ‘reference costs’ (second column) are the values we have used in our calculations, they lie in the middle of the range cited in the literature (third column). These ranges are broad because there can be significant variabilities in cost, depending on the capture technology, the availability and price of steam and electricity, the size of the facility, and the concentration of CO₂ in the exhaust flue gas.

\subsection{Transport and Storage}
\label{subsec:transport}
Since the aim of this work is to assess the immediate potential for negative emissions in Sweden, as a representative case, we have considered CO₂ storage at the \textit{Sleipner T} site, located in the Norwegian parts of the North Sea, which is the facility that is most readily available and furthest from Swedish point sources. This means that our transport costs can be seen as an upper limit. The CO₂ storage potentials in Sweden and neighbouring countries are described in~\ref{app:storage}.

Storage costs at the Sleipner facility have been well described~\cite{Chadwick2008}. These costs can be broken down into capital expenditure of almost 100~M€ and operational cost of 7~M€ per year. The largest proportion of the costs relates to the compressor. Considering that 16~Mt CO₂ have been injected at Sleipner over the last 21 years, it can be safely assumed that 15€/t CO₂ is the upper limit of the injection costs at this facility. This is in good agreement with the range of 6-20 €/t reported previously~\cite{ZEP2011b} for offshore storage in saline aquifers, which is the most expensive storage option.

Concerning transportation, it is clear that for Sweden to deploy BECCS on a large scale, it would be reasonable to design networks in which hubs would collect the CO₂ of several facilities before shipping to Sleipner. However, it is beyond the scope of the present work to try to design such CO₂ networks across Sweden (see~\cite{Kjarstad2016a}, for such analysis). Instead, we simply wish to provide an estimate of the transportation cost. The facility that is located furthest from Sleipner is on the border with Finland, near Haparanda, that is approximately 2500~km away by sea from Sleipner. Computing all the transportation costs with this value of 2500~km, our cost estimates remain highly conservative. Based on a previous work~\cite{ZEP2011}, we have used of the results concerning so-called "Network 2" which consists of 10~km of onshore pipelines followed by ship transportation for various distances, for a CO₂ quantity of 2.5~Mt/y. Extrapolating their values to 2500~km gives a cost of 27€/t which is in approximate agreement with the previously reported outcomes~\cite{Kjarstad2016a}. According to \cite{ZEP2011}, a cost variation of $\pm$50\% can be expected.

These numbers are summarized in Table~\ref{tab:trStCost}.

\begin{table}[ht]
\caption{\label{tab:trStCost}Costs related to CO₂ transportation and storage.}
\begin{indented}
\item[]\begin{tabular}{@{}lccc}
\br
Activity & Reference cost [€/t] & Optimist cost [€/t] & Conservative cost [€/t] \\
\mr
Transport & 27 & 13.5  & 40.5 \\
Storage & 15 & 6 & 20\\
\br
\end{tabular}
\end{indented}
\end{table}

Note that compression costs are included in both the transport and storage costs, which means that in an actual system, the total costs would probably be lower.

\subsection{Capture and Storage rate}
\label{subsec:rates}
The capture, transport, and storage of CO₂ cannot be accomplished at 100\% efficiency and every step is accompanied by some release of CO₂. In this work, following~\cite{Onarheim2015}, we assumed a CCS-rate of 85\% for the overall cycle. This excludes life cycle emissions linked to power consumption and transportation.

\subsection{Other considerations}
Overall costs are heavily dependent upon the specific facility and the timing of construction and operation and include varying costs that reflect of the state of the economy, price of electricity, required rates of return on capital, choice of technology, and project execution. Other factors outside the scope of this analysis include dynamic feedback from increased electricity use in the Nordic electric system, choice of transport fuels (which may be biofuels produced in BECCS facilities), cost savings from scaling and network effects, and the global impacts of opening up negative emissions pathways.

\section{Results}

\begin{figure}[ht]
\begin{center}
\includegraphics[width=0.6\textwidth]{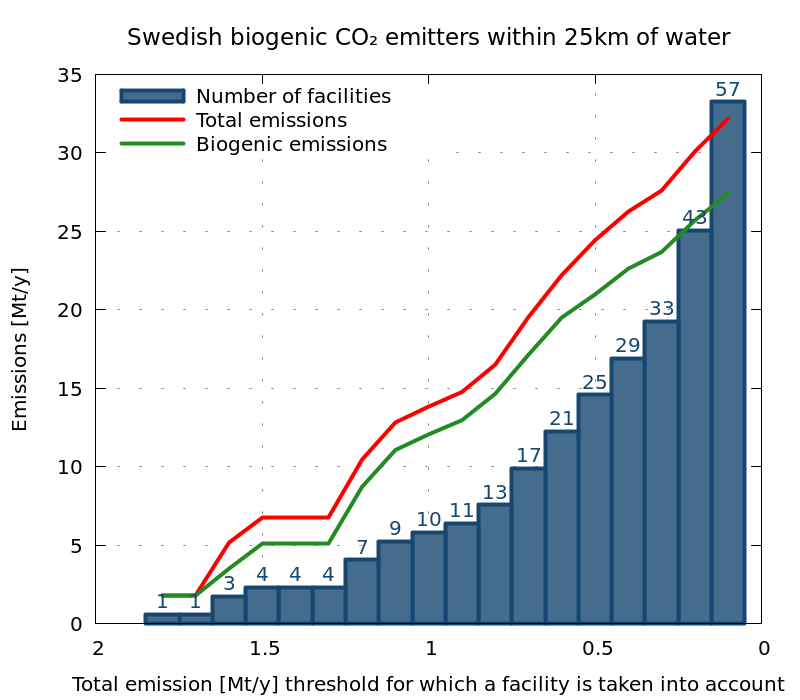}
\caption{Total CO₂ emissions (red), biogenic CO₂ emissions (green) and cumulative number of facilities (blue, labels above box diagrams) for the Swedish facilities that have emissions levels above the threshold on the \textit{x}-axis, have non-zero biogenic CO₂ emissions, and are located less than 25~km from the sea or Lakes Vänern and Mälaren.}
\label{fig:deployment}
\end{center}
\end{figure}

Applying the constraints detailed in section~\ref{subsec:selection} to the 96 large Swedish point sources of the E-PRTR results in 33 facilities that cumulatively account for 27.6~Mt CO₂, of which 23.7~Mt CO₂ are from biogenic origin, as can be seen in Figure~\ref{fig:deployment}. Applying the 85\% capture rate of \ref{subsec:rates}, this results in 23.5~Mt CO₂ stored, of which 20.1~Mt CO₂ are from biogenic origin. These 20.1~Mt CO₂ represent the immediate potential for negative emissions in Sweden.  
This corresponds to an offset of 39\% of current GHG emissions in Sweden. In addition, CCS at these facilities would reduce fossil CO₂ emissions by 3.4~Mt, which corresponds to 7\% of current Swedish GHG emissions. 

In Figure~\ref{fig:deployment}, it can be seen that applying CCS to the remaining 24 biogenic Swedish point sources (those having total CO₂ emissions <300 kt/y) would allow the generation of an extra 3.1~Mt/y of negative emissions, and reduce fossil emissions by an additional 0.7~Mt/y. This corresponds to an additional offset of 6\% of current Swedish GHG emissions, and a supplementary reduction of fossil emissions corresponding to 1\% of current Swedish GHG emissions.

It should be noted that CCS could also be applied at the 39 other Swedish point sources, which are located more than 25~km from the sea and/or are 100\% fossil fuel-powered. This could lead to 3.3 Mt/y of negative emissions and 9.9 Mt/y of fossil CO₂ emission reduction.

\begin{table}[ht]
\caption{\label{tab:SummaryCost}Costs related to CO₂ capture transportation and storage.}
\begin{indented}
\item[]\begin{tabular}{@{}lccc}
\br
Activity & & Reference cost [€/t] & Cost range [€/t]\\
\mr
Capture & Cement \& lime& 75 &40-110  \\
Capture & Pulp \& paper & 40 &16-62 \\
Capture & Power & 62 & 29.9-109.7\\
Capture & Ethanol fermentation & 25 & 18 - 40 \\
Transport & & 27 & 13.5 - 40.5 \\
Storage & & 15 & 6 - 20\\
\br
\end{tabular}
\end{indented}
\end{table}

The cost for capture, transport and storage applied in this report are summarized in Table~\ref{tab:SummaryCost}.

\section{Carbon price}

The current CO₂ tax in Sweden is 1.12 SEK per kg of CO₂~\cite{Wijkman2016,Skatteverket}, which is $\sim$120 €/t CO₂. While this represents one of the highest prices for CO₂ in the world, it should be noted that there are many tax exemptions and that this rate only applies to fuel for road transportation, industrial machines, and heating. Maritime transport, waste management, and agriculture are not liable for any kind of carbon tax~\cite{Wijkman2016}. As for the heating sector, the tax has had a real effect, causing the fuel used in the Swedish heating sector to be switched from fossil fuels (dominant in the 1970’s) to biomass (used predominantly today). While this is generally regarded as a strong example of the potency of a clear policy measure, its success is due in part to a combination of widespread adoption of district heating, the oil crises in the 1970’s, and a well-developed forestry industry, from which forest residues are available as fuel.

Most large industries are also exempted from the carbon tax because they are included in the European Emission Trading System (EU-ETS). The price of emission allowances under the EU-ETS, \textit{i.e.} the carbon price, has so far been significantly lower than the Swedish carbon tax ($\lesssim$~5€/t as of April 2017 \url{https://www.eex.com/}) and many industries receive large amounts of free credits, so far often even exceeding their actual emissions~(see~\cite{EEA}).

In Sweden, there are also an electricity certificate system, biofuel incentives and climate mitigation program tools (\textit{e.g.}, \textit{Klimatklivet}), as well as a bonus-malus system for new car sales, all of which are likely to have even higher implicit carbon costs.

\begin{figure}[ht]
\begin{center}
\includegraphics[width=0.7\textwidth]{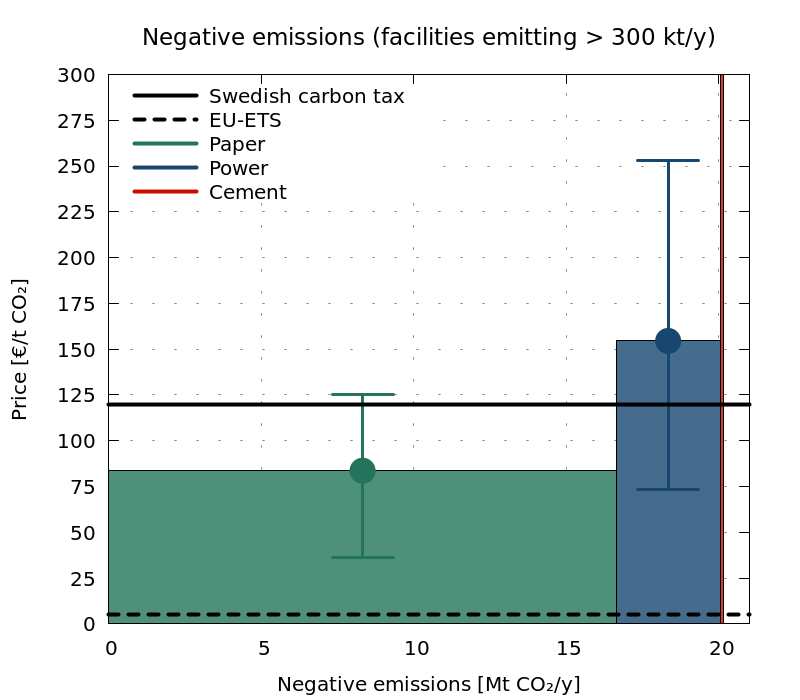}
\caption{Cost of negative emissions with BECCS [in €/t$_{CO_2}$] as a function of the  cumulative amount of negative emission readily achievable. Green corresponds to paper and pulp industry, blue to power and heat generation and red to cement.
The dots in the middle of the error bars are the respective costs when the reference values of Tables\ref{tab:captcost} and \ref{tab:trStCost} are used. The extent of the error bars represent the cost ranges from the same tables. As a reference the price of the Swedish carbon tax is displayed with a solid black line, the price of the EU-ETS credits as a dashed line.}
\label{fig:socben}
\end{center}
\end{figure}

As is evident from Figure~\ref{fig:socben}, since the price of the EU-ETS credit is only around 5 €/t$_{CO_2}$, industries have little incentive to lower their emissions. But for the paper and pulp industry, negative emission costs are lower than the Swedish carbon tax. If incentive mechanisms were to be slightly modified, industries could sell their negative emission carbon offsets to those that are paying the tax, for instance transportation or logistics companies.

Also from Figure~\ref{fig:socben}, it is clear that implementing BECCS at the Swedish paper industry would create large benefits. If the capture costs were to be kept at the lower end of the current cost estimate range, the same would go for the electricity generation sector. However, the Swedish cement industry does not emit enough biogenic CO₂ to generate negative emissions at an interesting price. If one considers only those technologies for which the cost for negative emissions would be lower than the carbon tax, then BECCS deployment would yield 16.7~Mt/y of negative emissions, giving rise to an annual cost saving (compared to current incentive marginal abatement costs) of 604~M€. 

As shown in Table~\ref{tab:negemi}, reaching the lowest end of the CCS cost range would allow to make the power production sector generate affordable negative emissions with respect to the Swedish carbon tax. We have not considered the case in which biomass would be further used, although of course this could lead to higher levels of negative emissions and cost savings.

\begin{table}[ht]
\caption{\label{tab:negemi}Negative emissions potential and marginal abatement costs with respect to the Swedish carbon tax.}
\begin{indented}
\item[]\begin{tabular}{@{}lcc}
\br
Scenario & negative emissions [Mt/y] & cost savings [M€/y] \\
\mr
Reference CCS costs / carbon tax & 16.7 & 604  \\
Lowest CCS costs / carbon tax  & 20.0 & 1 552\\
\br
\end{tabular}
\end{indented}
\end{table}

\section{Discussion}
We have seen here that even without increasing its use of biomass, Sweden already has the potential to create substantial negative emissions corresponding to 32\% of its total annual GHG emissions. BECCS has the potential to facilitate Sweden in becoming the first carbon-neutral welfare country, in line with current political ambitions. Negative emissions provide an opportunity to meet targets in a more economically feasible way and at an earlier time-point and/or to meet more ambitious targets, such as net-negative emissions at the national level. In fact, if the policies designed to cut GHG emissions by 85\% before Year 2045 are indeed implemented as promised, Sweden could become the first country in the world with a negative carbon balance, thereby starting the process of paying off its historical emissions debt. In light of the work of~\cite{Rockstrom2017}, which shows that the world needs to achieve 100–500 Mt CO₂/y of negative emissions by the mid-2030’s, we can say that Sweden has the potential to make an important contribution to this goal by producing around 17 Mt CO₂/y of negative emissions even ahead of this date. This would make Sweden an important international role model for others to follow, and if successful would contribute to innovations and new knowledge.

Our present study shows that economic incentives are already at the level required for negative emissions with BECCS to be a competitive mitigation technology. Nonetheless, changes to policy measures and regulatory frameworks are required to encourage the pulp and combined heat and power industries associated with the biogenic emission sources to act to realize large-scale implementation of BECCS.

\ack
This work is supported by the Swedish Innovation Agency (Vinnova) through the PAMOKO project in the \textit{Innovationer för ett hållbart samhälle} program (Diarienummer 2016-03387). The authors would like to thank the Global Lake Ecological Observatory Network (GLEON) and especially Dr. Hilary Dugan for making the shapefiles of Lake Vänern available online.

\section*{References}

\bibliography{article}

\appendix

\section{Estimating the distance from a facility to the sea}
\label{App:distances}
Since our study targets only the “low-hanging fruits”, we have identified which of the Swedish facilities are close enough to the sea to allow CO₂ transport by boat, which is the least-cost option~\cite{Kjarstad2016a}. To estimate the distances between the facilities and the sea, we have taken the geographical coordinates of the facilities (available in the E-PRTR) and calculated their distances to the nearest shoreline. The shoreline coordinates are derived from the shapefiles made available by Natural Earth\footnote{\url{naturalearthdata.com}}, \verb"ne_10m_coastline", which contains the coastlines of the entire world. Distances were calculated making use of the \verb"gDistance()" routine of the \verb"Rgeos" package~\cite{Bivand2017} in \verb"R"~\cite{RCoreTeam2015}. To allow  sufficiently accurate distance estimates, all the coordinates have been projected on the Coordinate Reference System EPSG:3035 which is highly suitable for Europe~\footnote{See \url{http://spatialreference.org/ref/epsg/3035/}}.

Besides sea transportation, Sweden has several large lakes connected to the sea by different canals which enable ship transportation from lake harbors through canals to the open sea. 
\begin{itemize}
\item Lake Vänern, which is the largest lake in the EU (area, 5650~km²), is connected to the Kattegat Sea \textit{via} the Trollhätte canal which goes from Vänersborg to the Göta älv river, which enters the sea at Gothenburg. Lake Vänern hosts several pulp and paper plants as well as power plants fueled by biomass.
\item Lake Vättern is the second largest lake in Sweden. It also has several biogenic CO₂ emitters located on its shores. Lake Vättern is connected to both the Baltic Sea and Lake Vänern through the Göta canal, which is quite shallow and narrow.
\item Lake Mälaren is the third largest lake in Sweden and concentrates a significant fraction of the country's economic activities. It is connected to the Baltic Sea through the Södertälje canal.
\item Lake Hjälmaren is the fourth largest lake in Sweden but is much smaller (483~km² or half the size of Lake Mälaren). It is connected to Lake Mälaren by the narrow Hjälmare canal.
\end{itemize}
None of the remaining 10 largest lakes in Sweden are connected to the sea.

Shapefile maps of Vänern are available online~\footnote{\url{https://github.com/GLEON/SOS/tree/master/GIS/VanernLake/Vanern_Shapefile}}. The other three lakes have been cropped out of the Natural Earth shapefile, \verb"ne_10m_lakes" which contains most of major lakes of the World.

\section{Storage sites}
\label{app:storage}
As shown by~\cite{Anthonsen2014}, there are several geologic sites in the Nordic region that appear to be suitable for carbon storage. This also includes sites in Sweden~\cite{Mortensen2014}: Faludden, which lies south-east of Gotland in the Baltic Sea; and Arnager Greensand and Höganäs-Rya, both of which are located between Denmark and Sweden. Note that the Faludden formation extends into Danish, Polish, Lithuanian, Latvian and Russian territories which could make its exploitation administratively complex. Moreover, none of these Swedish sites have yet been used for CO₂ storage and more exploration will probably be required before any injection there can be considered.
However, one site in the Norwegian Sea, Sleipner T (58\degree ~22'~6.98''~N, 1\degree ~54'~23.33''~E)\footnote{See the webpage of the Norwegian Petroleum Directorate \url{https://goo.gl/qLtRq6}} is already used for CO₂ storage and since 1996, 16~Mt CO₂ have been injected in the Utsira formation injection site from the Sleipner T facility\footnote{See Statoil webpage \url{https://goo.gl/FjzJHq}}. According to~\cite{Chadwick2008}, the Utsira formation has a storage capacity of 270~Mt CO₂. Note however that this corresponds to the capacity directly that is accessible to the current well. According to more recent estimates~\cite{Halland2011}, the whole Utsira formation, together with the Skade formation that lies just underneath, could store up to 16~Gt$_{CO_2}$. It should also be noted here that injecting approximately 20~Mt CO₂/y at Sleipner would require implementation of water production, the costs for which have not been taken into account in this work. The fact that Sleipner T is already operational is the reason that we have chosen it as our reference case. Currently a new injection site is under development at the Smeaheia formation led by Gassnova, and this may become the most important injection site for CCS in the Nordic region. This formation is not much further from Sweden than Sleipner, so transferring the CO₂ storage there should not affect the transport costs of~\ref{subsec:transport} and could potentially lower the storage costs.

As pointed out by~\cite{Bjornsen2012}, there exist some concerns that the London Convention on the Prevention of Marine Pollution by Dumping of Wastes and Other Materials (LC'72) does not allow for the export of  CO₂ streams from one country to another. The guidelines adopted in 2012~\cite{LC72} should resolve this issue provided that a sufficient number (29) of parties sign the Year 2009 amendment to the LC'72 that allows for the storage and export of CO₂. While this is a lengthy process, there may be room for alternative interpretations of the existing amendment currently under debate~\cite{Henriksen2016}.

\section{Swedish facilities}
\label{app:prtr}

In the E-PRTR, countries are obliged to report the total CO₂ emissions from their facilities and have the \textit{possibility} to report "non-biogenic" CO₂ emissions if they so wish. The absence of mandated reporting for non-biogenic CO₂ emissions for a facility may mean that the country does not report this quantity or that the facility only emits biogenic CO₂, or in some cases that it emits only non-biogenic CO₂, which prevents any automatic exploitation of the data. The situation is further complicated by the fact that several countries (Bulgaria, Germany, Romania, Slovenia and The Netherlands) appear to report non-biogenic emissions for some of their facilities only. As a consequence, it is not possible to know whether the facility is 100\% biogenic or just simply not reporting non-biogenic CO₂ emissions.

There are three Swedish facilities in the E-PRTR for which there are no reports concerning non-biogenic CO₂ emissions. From the Swedish Pollutant Release and Transfer Register\footnote{\url{http://utslappisiffror.naturvardsverket.se}} of the Swedish Environmental Protection Agency (\textit{Naturvårdsverket}), we discovered that these three facilities are indeed emitting 100\% biogenic CO₂.

In the E-PRTR, each facility is reported with two fields concerning its activities and the information provided is sometimes conflicting. For instance the Hedensbyn kraftvärmeverket is registered under both \textit{Manufacture of other products of wood; manufacture of articles of cork, straw and plaiting} (NACE) and \textit{Thermal power stations and other combustion installations} (IA). The likely reason for this is that this power and heat plant also produces wood pellets. Likewise, the number of factories registered under \textit{Manufacture of Pulp} and \textit{Manufacture of Paper} are not the same in the NACE and the IA classifications, which is probably because some of them are integrated pulp and paper plants. This justifies our regrouping of all the facilities into the four categories of \ref{tab:captcost} only.

\begin{table}[ht]
\caption{\label{tab:data}Swedish facilities emitting biogenic CO₂ and located less than 25~km from the sea, or lakes Vänern and Mälaren.}
\begin{indented}
\item[]
\begin{tabular}{@{}cllcc}
&\bfseries Name & \bfseries Activity & \bfseries Total & \bfseries Biogenic\\
&& & \bfseries CO₂ [kt/y]& \bfseries CO₂ [kt/y]\\ \hline
    \begin{filecontents*}{Sweden.csv}
,FacilityName,Activity,Distance to coast(km),Total CO₂ (kt),Total biogenic CO₂ (ktonne)
1,Cementa Degerhamn,Cement,0,259,4
2,Cementa AB Slitefabriken,Cement,0,1700,120
3,Lantmännen Agroetanol AB,Ethanol,1,140,138.58
4,BillerudKorsnäs Karlsborgs AB,Paper,0,791,781.98
5,SCA Ortviken,Paper,0,339,316.6
6,Korsnäsverken,Paper,0,1250,1236.2
7,SCA Obbola AB,Paper,0,484,455.6
8,Iggesunds Bruk,Paper,0,864,853.8
9,Skoghalls Bruk,Paper,0,878,828.1
10,Gruvöns bruk,Paper,0,1220,1196.8
11,Bravikens Pappersbruk,Paper,0,148,138.16
12,Mondi Dynäs AB,Paper,1,564,555.21
13,SCA Munksund,Paper,1,688,654.7
14,Stora Enso Nymölla AB,Paper,1,685,685
15,Smurfit Kappa Kraftliner Piteå,Paper,1,1190,1180.04
16,Metsä Board Sverige AB,Paper,1,1650,1601.1
17,Bäckhammars Bruk,Paper,7,462,454.58
18,BillerudKorsnäs Sweden AB,Paper,17,1010,998.2
19,Munksjö Paper AB Billingsfors,Paper,23,179,162.5
20,SCA Östrands massafabrik,Paper,0,1170,1156.8
21,Vallviks Bruk,Paper,0,608,602.03
22,Skutskärs Bruk,Paper,1,1590,1579.4
23,Södra Cell Värö,Paper,1,1230,1204.7
24,Södra Cell Mönsterås,Paper,1,1830,1801
25,Södra Cell Mörrum,Paper,1,930,900.5
26,Domsjö Fabriker AB,Paper,3,528,525.51
\end{filecontents*}
\csvreader[]{Sweden.csv}{}
    {\\\csvcoli&\csvcolii&\csvcoliii&\csvcolv&\csvcolvi}
\end{tabular}
\end{indented}
\end{table}

\begin{table}[ht]
\caption{\label{tab:data2}Same as Table~\ref{tab:data} (end).}
\begin{indented}
\item[]
\begin{tabular}{@{}cllcc}
&\bfseries Name & \bfseries Activity & \bfseries Total & \bfseries Biogenic\\
&& & \bfseries CO₂ [kt/y]& \bfseries CO₂ [kt/y]\\ \hline
    \begin{filecontents*}{Sweden2.csv}
,FacilityName,Activity,Distance to coast(km),Total CO₂ (kt),Total biogenic CO₂ (ktonne)
27,Hedensbyns kraftvärmemerk,Power,4,213,166.7
28,Västerås kraftvärmeverk,Power,0,688,426
29,Filborna Kraftvärmeverk,Power,5,248,186
30,Hässelbyverket,Power,0,276,272.01
31,Lidköpings Värmeverk Filen,Power,0,268,218.9
32,Hedenverket,Power,0,253,234.9
33,Solnaverket,Power,0,121,119.1
34,Bomhus Energi,Power,0,255,255
35,Västhamnsverket (VHV),Power,1,120,119.39
36,Sundsvall Energi AB Korstaver,Power,1,178,114.5
37,Igelsta kraftvärmeverk,Power,1,750,724.8
38,Igelsta värmeverk,Power,1,337,212
39,Händelöverket,Power,1,731,535
40,Värtaverket,Power,1,595,77
41,HPC Simpan och Ena Kraft kraftvärmeverket,Power,2,133,130.1
42,Idbäckens Kraftvärmeverk,Power,2,158,157.1
43,Kraftvärmeverket Övik Energi A,Power,3,256,226.1
44,Bristaverket,Power,4,433,372.7
45,Riskullaverket,Power,5,172,163.55
46,Boländer: Avfallsförbränning,Power,6,348,216
47,Boländer: Kraftvärmeverket,Power,6,192,43
48,Jordbro kraftvärmeverket,Power,6,251,251
49,Högdalenverket,Power,6,767,451
50,Sävenäs Kraftvärmeverk,Power,7,138,132.05
51,Dåva kraftvärmeverket,Power,7,444,357.2
52,Sävenäs,Power,8,532,313
53,Värmeverket Vattumannen,Power,8,319,317.33
54,Lillesjö Avfallskraftvärmeverk,Power,8,108,65.4
55,Bodens värmeverk BEAB,Power,12,124,85.3
56,Allöverket,Power,16,202,201.87
57,Kristinehedsverket,Power,5,195,141.4
\end{filecontents*}
\csvreader[]{Sweden2.csv}{}
    {\\\csvcoli&\csvcolii&\csvcoliii&\csvcolv&\csvcolvi}
\end{tabular}
\end{indented}
\end{table}

\end{document}